\documentclass[aps,prb,twocolumn]{revtex4-1}
\usepackage[colorlinks=true,linkcolor=blue,urlcolor=blue,citecolor=blue,pdfusetitle]{hyperref}
\usepackage[utf8]{inputenc}
\usepackage[english]{babel}
\usepackage[normalem]{ulem}
\usepackage{amsmath}
\usepackage[caption = false]{subfig}
\usepackage{graphicx,epstopdf}
\usepackage{blindtext}
\usepackage[table,xcdraw]{xcolor}
\usepackage{lipsum}
\usepackage{amsfonts}
\usepackage{bbm}
\usepackage{xfrac}
\usepackage{amssymb}
\usepackage{enumerate}
\usepackage{color}
\usepackage{latexsym}
\usepackage{physics}
\usepackage{times,txfonts}
\usepackage{tikz}
\usetikzlibrary{quantikz2}
\usepackage{graphicx}

\newcommand{\1}{\mathbbm{1}}

\definecolor{MyGreen}{RGB}{0, 179, 134}
\definecolor{MyRed}{RGB}{255, 102, 102}

\newcommand{\UFF}{Instituto de F\'{i}sica, Universidade Federal Fluminense, Av. Gal. Milton Tavares de Souza s/n, Gragoat\'{a}, 24210-346 Niter\'{o}i, Rio de Janeiro, Brazil}

\usepackage{orcidlink}

\begin{document}

\title{Variational simulation of quantum phase transitions induced by boundary fields}

\author{Alan Duriez~\orcidlink{0000-0001-5196-0827}}
\email{alanduriez@id.uff.br}
\affiliation{\UFF}

\author{Andreia Saguia~\orcidlink{0000-0003-0403-4358}}
\email{asaguia@id.uff.br}
\affiliation{\UFF}

\author{Marcelo S. Sarandy~\orcidlink{0000-0003-0910-4407}}
\email{msarandy@id.uff.br}
\affiliation{\UFF}

\begin{abstract}
     The characterization of quantum phase transitions is a fundamental task for the understanding of quantum phases of matter, with a number of potential applications in quantum technologies. In this work, we use digital quantum simulation as a resource to theoretically and experimentally study quantum phase transitions. More specifically, we implement the variational quantum eigensolver (VQE) algorithm to the one-dimensional spin-$1/2$ transverse-field Ising chain in the presence of boundary magnetic fields. Such fields can induce a rich phase diagram, including a first-order line and also a continuous wetting transition, which is a quantum version of the classical wetting surface phenomenon. We present results for noiseless simulations of the associated quantum circuits as well as hardware results taken from a superconducting quantum processor. For different regions of the phase diagram, the quantum algorithm allows us to predict the critical value of the magnetic fields responsible for either the first or second-order transitions occurring in the system. 
     
\end{abstract}

\maketitle

\section{Introduction}

As envisaged by Feynman~\cite{Feynman:82}, quantum computers may be an efficient resource to simulate many-body physics, which is a hard task for classical computers even at an intermediate scale (tens to hundreds of particles). In this scenario, quantum processors in the pre-fault tolerant era emerge as candidates to achieve a possible quantum advantage, particularly when used for the simulation of condensed matter physics. The quest for a quantum advantage on current quantum devices has motivated the search for algorithms that are adapted to sparse-qubit connectivity, short decoherence times, and large amounts of noise, both in measurement and in gate operations. As a recent approach to this problem, variational quantum algorithms~\cite{Cerezo:2021} aim at reducing the expenses of quantum resources by including a classical optimization routine applied to a parametrized quantum circuit. In particular, an interesting proposal for physical applications is the variational quantum eigensolver (VQE) \cite{Tilly:2022}, which is suitable for finding eigenstates of highly correlated systems described in quantum chemistry, many-body physics, and condensed matter applications. Many different platforms have been used to simulate quantum systems with VQE, including trapped ions \cite{Zhu:2020,Pagano:2020}, photonics \cite{Lee:22,Yao:2022}, and superconducting qubits~\cite{Kandala:2017,Andi:2021,Grossi:2023,Han:2023}.

In this work, we show how we can apply the VQE algorithm for the analysis of spin chains undergoing quantum phase transitions (QPTs) induced by boundary fields. Such fields may generate domain walls in the system, leading to the formation of magnetic domains of aligned spins. These domains are separated by localized disturbances (defects) called kinks, which appear for critical values of anti-parallel boundary fields. Quantum criticality can be characterized by non-analyticities in the properties of the ground state of the system, with  this behavior observed by simulating the ground state of the system in different points across the phase space. For example, first and second-order QPTs are present in a variety of many-body systems~\cite{Piazza:99,Uhlarz:04,Knafo:09}. They can be identified by discontinuities in the first derivative and divergences in the second derivative of the ground state energy with respect to a relevant driving parameter, respectively~\cite{Sachdev:Book}. Here, we apply VQE to simulate the one-dimensional spin-$1/2$ transverse-field Ising chain in the presence of an anti-parallel boundary magnetic field. For this model, there is a second-order critical boundary field that separates a ferromagnetic phase from a kink-type phase~\cite{Campostrini:14,Campostrini:15}. The ferromagnetic phase is characterized by an exponentially decreasing ground state energy gap, {\it i.e.}, $\Delta_L \sim e^{-cL}$, with $c$ a constant factor and $L$ denoting the length of the chain. For the kink-type phase, we have a polynomial decay, {\it i.e.}, $\Delta_L \sim 1/L^{p}$, with $p$ a constant factor. The emergence of the kink-type phase turns out to be associated with a sharp increase in the expectation value of the kink operator. In particular, this kink-type phase can be interpreted as a quantum version of the so-called wetting phenomenon~\cite{Hu:2021}, which constitutes an active field of research in phase transitions~\cite{Wetting-RMP:09}. As we will show, quantum wetting can be simulated by a VQE quantum circuit, which in turn can be physically realized in a quantum processing device. By using the IBM Torino quantum processor, we will then experimentally implement the quantum simulation of the magnet-to-kink QPT present in the quantum phase diagram. 

We will begin by exactly diagonalizing the Hamiltonian, identifying the QPT by looking at the ground state energy as a function of the boundary field. Next, we will present the results obtained via VQE, where we will show that it is possible to match the exact results with high accuracy, while considering noiseless simulations of the associated quantum circuits. Moreover, we show how we can prepare the ground state of the system on a quantum processor and extract meaningful physical properties that are in close agreement with the exact results, even in the presence of hardware noise. The quantum circuits for VQE are built using the Qiskit SDK~\cite{qiskit2024} and performed experimentally on the IBM quantum infrastructure.  

\section{Results}

The physical system we consider is the 1D spin$-1/2$ transverse-field Ising model with open boundary conditions under anti-parallel magnetic fields applied at the borders (first and last sites), in the same direction as the coupling axis. The Hamiltonian for this system can be written as
\begin{figure*}[t] 
    \centering
    \includegraphics[scale=0.98]{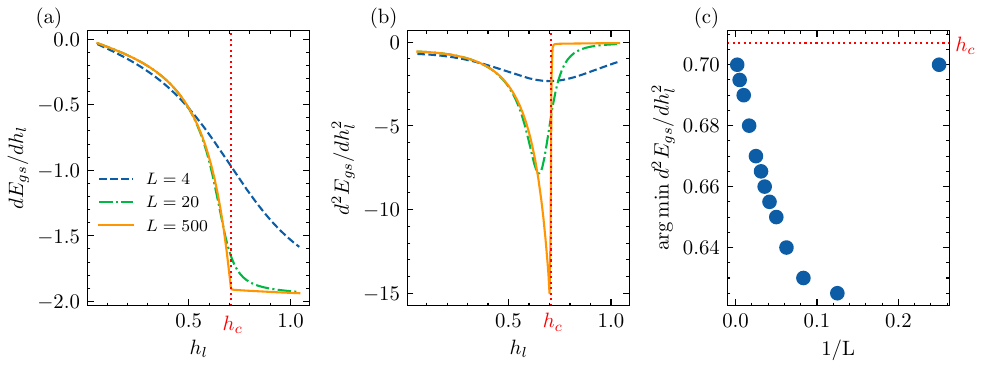} 
    \caption{{\bf (a)} Exact first derivative of the ground-state energy of $\hat{H}$ for different sizes $L$ with respect to the boundary field $h_l$, obtained from the analytical solution, considering $h_x=0.5$, $J=1$, and $h_r=-h_l$. 
    {\bf (b)} Exact second derivative of the ground-state energy of $\hat{H}$, with the same parametrization as in (a). 
    {\bf (c)} The argument of the minimum of the second derivative of the ground-state energy as a function of $1/L$ for $L=\{4,8,12,16,20,24,28,32,40,60,100,200,500 \}$. We also indicate the critical value of the boundary field $h_c=\sqrt{1-h_x}\approx0.71$ with a red dotted line. Notice the scaling of the precursor of quantum critical point towards the asymptotic critical value as  $L\rightarrow\infty$. The plots are in units such that $J=1$.}
    \label{fig:jordan_wigner_magkink}
\end{figure*}
\begin{align}
    \hat{H} = -J\sum_{i=1}^{L-1}\hat{\sigma}^{(z)}_i \hat{\sigma}^{(z)}_{i+1} -h_x\sum_{i=1}^{L}\hat{\sigma}^{(x)}_i+h_l \hat{\sigma}^{(z)}_1+h_r \hat{\sigma}^{(z)}_L \label{eq:hamiltonian}
\end{align}
where $J$ and $h_x$ are the coupling and transverse-field parameters, respectively, and $h_l$ and $h_r$ are the left and right boundary fields intensities, respectively, with $\hat{\sigma}^{(z)}_i$ and $\hat{\sigma}^{(x)}_i$ denoting the Pauli matrices at site $i$, and $L$ is the length of the spin chain. Throughout this work, we take $J=1$ and assume  $0<h_x<1$. We also consider the boundary fields to be in opposite directions, therefore adding the constraint $h_l\:h_r < 0$. This system undergoes different types of phase transitions as we change the values of the boundary field while keeping the coupling and transverse field fixed. These phase transitions caused by boundary fields were first described by Ref.~\cite{Campostrini:15}. In this work, the authors restricted the analysis to the case where the boundary fields had the same intensities but opposite directions, or $h_r=-h_l$. They showed that the system transitioned from a magnetic phase to a kink-type phase at the critical anti-parallel boundary field $h_c=\sqrt{1-h_x}$, with striking differences in the energy spectrum for the two phases. This characterization is captured by the  energy gap $\Delta_L = E_L^{(1)} - E_L^{(0)}$ between the first-excited state, with energy $E_L^{(1)}$, 
and the ground state, with energy $E_L^{(0)}$. Specifically, the magnetic phase is characterized by an exponentially decreasing energy gap $\Delta_L \sim e^{-cL}$, with $c$ a constant factor. On the other hand, for the kink-type phase, we have a polynomial decay, {\it i.e.}, $\Delta_L \sim 1/L^{p}$, with $p$ a constant factor. Moreover, as we can show here, the emergence of the kink-type phase turns out to be 
associated with a sharp increase in the expectation value $\langle \hat{N}_k \rangle$ of the kink operator
\begin{equation}
    \hat{N}_k = \frac{1}{2}\sum_{i=1}^{L-1}
(\1 -\hat{\sigma}^{(z)}_i \hat{\sigma}^{(z)}_{i+1}). \label{eq:number-kinks}
\end{equation}
Here, we will denote this transition as the {\it magnet-to-kink} QPT. 
The boundary field quantum criticality has been further analyzed in Ref.~\cite{Hu:2021}, where a more complete picture of the phase diagram of these models is drawn. In particular, the Hamiltonian in Eq.~(\ref{eq:hamiltonian}) exhibits both first and second-order QPTs for different values of the boundary fields. Considering a two-dimensional phase diagram for the variables $h_l$ and $h_r$, while keeping $J=1$ and $0<h_x<1$ fixed (see Ref.~\cite{Hu:2021}), the segment associated with $-\sqrt{1-h_x}<h_r=-h_l<\sqrt{1-h_x}$ is the border of a first-order phase transition between the "Positive" and "Negative" phases ({\it positive-to-negative} QPT), following the nomenclature  in Ref.~\cite{Hu:2021}. Moreover, the lines $\left(h_l > \sqrt{1-h_x}, h_r = - \sqrt{1-h_x}\right)$ and $\left(h_l = \sqrt{1-h_x},h_r < - \sqrt{1-h_x}\right)$, as well as their symmetric reflection versions $\left(h_l < - \sqrt{1-h_x}, h_r =  \sqrt{1-h_x}\right)$ and $\left(h_l =  - \sqrt{1-h_x},h_r >  \sqrt{1-h_x}\right)$, are borders between the positive or negative phases and a wetting phase. This terminology comes from the well-known map between the 1D quantum Ising model and the 2D classical Ising model, which exhibits a continuous wetting-type transition that was exactly solved by Abraham in 1980~\cite{Abraham:1980}. We revisit the quantum phase diagram and the analytical solution of the model in Appendix~\ref{App_A}. 
From this picture, we can see that the magnet-to-kink transition discussed in Ref.~\cite{Campostrini:15} is a transition between the positive-to-negative critical line and the wetting region. We can verify both the order of the QPT and the critical points by showing the derivatives of the exact ground state energy as a function of the boundary fields. We can obtain these energies analytically using the Jordan-Wigner transformation, which maps the Hamiltonian shown in Eq.~(\ref{eq:hamiltonian}) into an effective model of free fermions~\cite{Hu:2021}. By adopting this procedure, we show in Figs.~\ref{fig:jordan_wigner_magkink}(a) and~\ref{fig:jordan_wigner_magkink}(b) the derivatives of the ground-state energy as a function of $h_l$, with $h_r=-h_l$ and $h_x$ fixed, for increasing values of $L$, so that the thermodynamic behavior can be analyzed. By inspecting the minimum in the second derivative of the ground-state energy, which tends to a non-analyticity as we increase the value of $L$, we can see that this particular transition is of second order. 
In Fig.~\ref{fig:jordan_wigner_magkink}(c), we show the  argument of the minimum of the second derivative as a function of $1/L$, which indicates how the critical point $\sqrt{1-h_x}$ is approached as $L\rightarrow\infty$. 
Notice that, for $L$ small (up to $L=8$ in the plot), the precursor of the critical point for finite $L$ moves away from the expected asymptotic limit, approaching it in the correct direction for sizes $L\ge 12$ in Fig.~\ref{fig:jordan_wigner_magkink}(c). This unusual scaling of the critical point for finite sizes can be seen as a further challenge posed by boundary fields, which require larger chains to achieve the asymptotic critical behavior. In Appendix~\ref{App_B}, we also explore the behavior of the first- and second-order QPTs induced by $h_l$, but obtained with the value of $h_r$ constant.

To obtain the ground state in different regions of the phase space we can also employ the Variational Quantum Eigensolver (VQE) algorithm. In VQE, we use a parametrized quantum circuit, or ansatz, to prepare trial states to approximate the ground state of the system, and we then optimize the parameters of the circuit in a quantum-classical feedback loop to search for the parameters that minimize the energy of the prepared state. For a given vector of parameters $\boldsymbol{\theta}$, the output state of the variational ansatz can be represented as
\begin{align}   \ket{\psi(\boldsymbol{\theta})}&=U(\boldsymbol{\theta}) U_I\ket{0}^{\otimes L}, \label{eq:ansatz}
\end{align}
where $U_I$ is the non-parametrized unitary associated with the initial state preparation, $U(\boldsymbol{\theta})$ is the parametrized sector of the ansatz, and $\ket{0}^{\otimes L}$ represents all qubits initialized in the computational state $\ket{0}$, with $\hat{\sigma}^{(z)}\ket{0}=\ket{0}$. 
To find the parameters for which the prepared state best approximates the true ground state of the system, we optimize them using a classical optimization algorithm in which the cost function is considered to be the expectation value of the energy evaluated on the state prepared by the ansatz, or 
\begin{equation}
    C(\boldsymbol{\theta})=\bra{\psi(\boldsymbol{\theta})} \hat{H} \ket{\psi(\boldsymbol{\theta})}. \label{eq:cost-function}
\end{equation}
For the first iteration of the feedback loop, the circuit is initialized with random parameters, denoted as $\boldsymbol{\theta}_0$, sampled from a uniform probability distribution. Next, the obtained value of the energy and the associated parameters are given as input to a classical optimization algorithm that evaluates the cost function at additional points in the vicinity of the starting point in the parameter space. With this additional evaluations, it is possible to predict a shift of the parameters $\delta\boldsymbol{\theta}$ for which the updated parameters $\boldsymbol{\theta}_1\leftarrow \boldsymbol{\theta_0}+\delta\boldsymbol{\theta}$ return a value of the cost function that best approximates the true ground-state energy of the system. This cycle is then repeated for $\boldsymbol{\theta}_1$, and so on, until some convergence criterion is achieved.

The success of the optimization will depend heavily on the properties of the cost function $C(\boldsymbol{\theta})$, and therefore on the structure of the parametrized circuit providing $\ket{\psi(\boldsymbol{\theta})}$ in Eq.~(\ref{eq:ansatz}). In this work, we consider the Hamiltonian variational ansatz (HVA) for circuit parametrization, which was first proposed in Ref.~\cite{Wecker:15} and further analyzed in Refs.~\cite{Wiersema:2020,Park:20} in the context of many-body quantum simulations. For a general Hamiltonian written as a sum of Pauli strings $\hat{H} = \sum_{\alpha}c_{\alpha} \hat{P}_\alpha$, the HVA ansatz with $p$ layers will be composed of rotations generated by the individual terms $\hat{P}_\alpha$. This circuit can generally be written as
\begin{align}   
\ket{\psi_\text{HVA}(\boldsymbol{\theta})}=\prod_{i=1}^p\prod_{\alpha}\exp{-i\theta_{i \alpha} \hat{P}_{\alpha}}\ket{\psi_0}, \label{eq:circuit-unitary}
\end{align}
where $\ket{\psi_0}$ is uniform superposition of computational basis states, and $\{\theta_{i\alpha}\}$ is the set of variational parameters, one for each rotation and each layer.

\begin{figure*}[t] 
    \centering
    \includegraphics[scale=0.98]{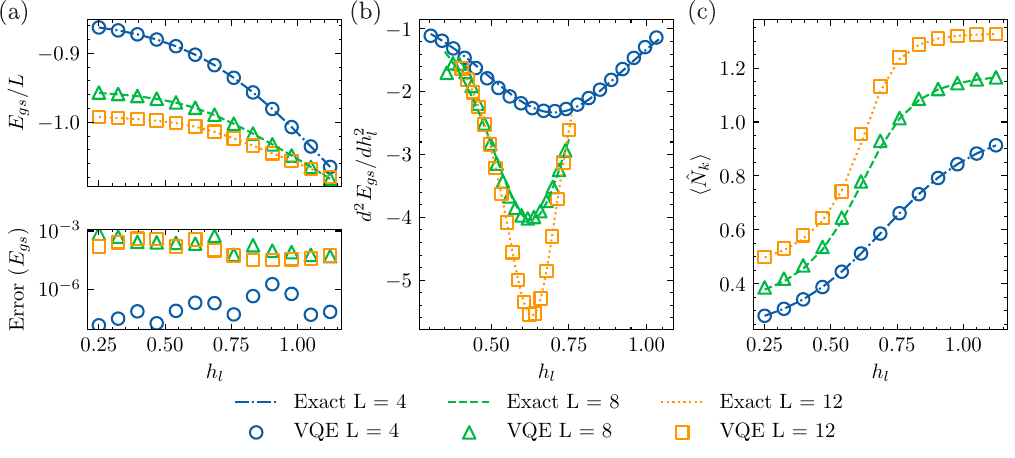} 
    \caption{VQE results obtained with noiseless simulations for the magnet-to-kink transition ($h_r=-h_l$) for $L=\{4,8,12\}$. We have considered $J=1$ and $h_x=0.5$, and $h_l$ and varied in the range $(0.4,1)$. These results were obtained with noiseless simulations of the associated quantum circuits. As a comparison to the VQE approach, we also exhibit the results obtained by exact diagonalization. {\bf (a)} Ground-state energy of $\hat{H}$, normalized by $L$, as a function of the boundary field $h_l$. In the bottom plot we have relative error of the energies in comparison with the exact energies $E_{\text{ex}}$ for the same values of the boundary field, which is given as $|(E_{ex}-E_{gs})/E_{ex}|$, with $E_{gs}$ being the ground state energy given by VQE. {\bf (b)} Second derivative of the ground-state energy as a function of $h_l$, obtained by interpolating the energy curve from the plot in (a). {\bf (c)} Expectation value of the kink operator as a function of $h_l$ evaluated for the ground state obtained from the VQE algorithm. The plots are in units such that $J=1$.}
    \label{fig:energy_magkink}
\end{figure*}

To obtain the position of the critical point and describe the character of the phase transition using VQE, we optimize the parameters of the ansatz while considering different values of the boundary field $h_l$ in a range that contains the critical point. The output for these ground-state energies will then be the value of the cost function as given by Eq.~(\ref{eq:cost-function}) evaluated for the optimized parameters. We then calculate the energy derivatives from the set of computed energy values, which can be done by standard numerical techniques. 
The VQE results obtained with noiseless simulations are  shown in Fig.~\ref{fig:energy_magkink}. 
In Figs.~\ref{fig:energy_magkink}(a) and~\ref{fig:energy_magkink}(b), we show the ground-state energy, as well as its second derivative, as a function of the boundary field $h_l$ for different values of $L$. In Fig.~\ref{fig:energy_magkink}(c), we show the behavior of the expectation value of the kink operator, $\hat{N}_k$, evaluated for the output state of VQE. As a comparison to the VQE results, the exact values have also been used for all the quantities plotted. They were obtained by simulating the considered quantum circuits on a classical computer, which accounts for computing the expectation values using the exact ansatz wavefunction. We can see that, for all sizes considered and for all values of $h_l$, the relative error on the ground state energy given by VQE is below $10^{-4}$. Moreover, the VQE predictions for the second derivative and the expectation value of the kink operator closely match the exact results. In Appendix~\ref{App_B}, we also include the VQE simulations for other types of transitions in different regions of the quantum phase diagram.

\begin{figure*}[htbp]
    \centering
    \includegraphics[width=0.47\textwidth]{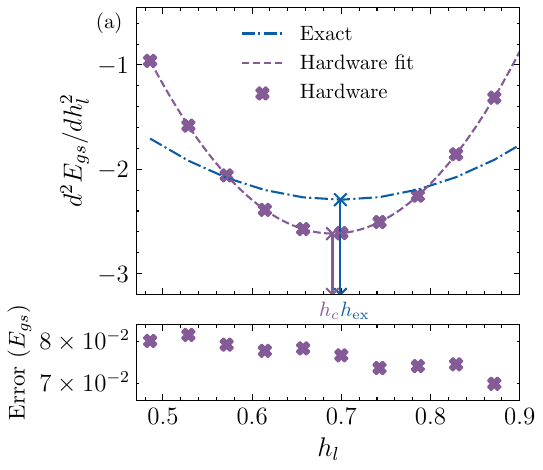}
    \hfill
    \includegraphics[width=0.47\textwidth]{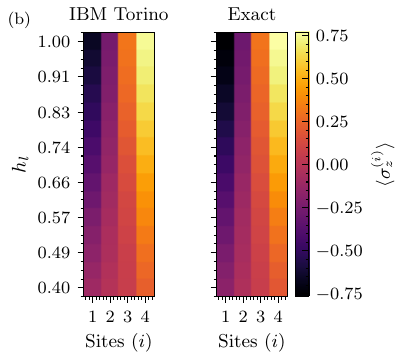}
    \caption{Quantum hardware results obtained from \texttt{ibm\_torino}. {\bf (a)} Second derivative of the ground state energy as a function of $h_l$ for the magnet-to-kink transition considering $L=4$, comparing the results with the data obtained from the quantum processor. We highlight the minimum of the curve obtained by interpolating the hardware data and the minimum of exact plot. By indicating the exact minimum as $h_{\text{ex}}$ and the one obtained from the hardware as $h_c$, the relative error in the prediction of the minimum is $|h_{\text{ex}}-h_c/h_{\text{ex}}|=0.013$. In the bottom of (a), we also show the relative error in the expectation value of the ground-state energy obtained on hardware, in comparison with the exact results. We can then see that the relative error in the prediction of the critical field $h_c$ is considerably smaller than the relative errors shown for the ground-state energy. This indicates that we may be able to probe the critical behavior for larger systems even in the presence of hardware noise. {\bf (b)} Expectation values of the local magnetization of the spins in the lattice for different values of the boundary field $h_l$, showing that we are able to obtain quantitatively accurate description of physical properties of the system using pre-fault tolerant quantum devices. The plots are in units such that $J=1$. }
    \label{fig:hardware_magkink_4q}
\end{figure*}

We can also obtain experimental results about the critical point with the VQE ground states prepared in a pre-fault tolerant quantum hardware device, namely, \texttt{ibm\_torino}. Indeed, in Fig.~\ref{fig:hardware_magkink_4q}(a), we show the second derivative of the ground-state energy, considering the magnet-to-kink transition for $L=4$. In Fig.~\ref{fig:hardware_magkink_4q}(b), we show the values of the local magnetization obtained from \texttt{ibm\_torino} compared with the exact magnetization. These results were obtained by computing $C(\boldsymbol{\theta}_{\text{opt}})$ for the ansatz state initialized with the optimal parameters $\boldsymbol{\theta}_{\text{opt}}$ obtained with the classical simulations of the VQE algorithm. 

By inspecting the ground-state energy plot in Fig.~\ref{fig:hardware_magkink_4q}(a), we can  see that, as a consequence of hardware noise, the expectation values of the second derivative of energy may differ from the exact values. However, the position of the minimum, which indicates the critical value of $h_l$, is only slightly affected by noise. An analysis of the energy error based on the root mean square (RMS) deviation and the effects on its second derivative is provided in Appendix~\ref{app_C}. By looking now at the magnetization plot in Fig.~\ref{fig:hardware_magkink_4q}(b), we can see that the noisy expectation values collected from the quantum device provide a correct description of the physical properties of the system, in particular, how the spins at the borders become magnetized in opposite directions as we increase the boundary fields $h_l$ and $h_r$. These results suggest that the VQE algorithm indeed allows for probing the critical behavior induced by the boundary fields in other pre-fault tolerant applications. 

\section{Methods}

\begin{figure}
\centering
\begin{quantikz}[column sep=0.42cm]
 & \gate{H}\slice{} & \gate{R_X(\theta_1)}  & \gate[2]{R_{ZZ}(\theta_2)} &  & \gate{R_Z(\theta_3)} & \\
 & \gate{H} & \gate{R_X(\theta_1)}  &  & \gate[2]{R_{ZZ}(\theta_2)} & & \\
 & \gate{H} & \gate{R_X(\theta_1)}  & \gate[2]{R_{ZZ}(\theta_2)} &  & & \\
 & \gate{H} & \gate{R_X(\theta_1)}  &                            &  & \gate{R_Z(-\theta_3)} &
\end{quantikz}
\caption{Circuit representation for the HVA ansatz used to simulate the ground state of $\hat{H}$ in the magnet-to-kink transition for $L=4$, considering a single layer $(p=1)$. The transverse and boundary fields are associated with the $R_X$ and $R_Z$ rotation gates, while the coupling term is associated with the $R_{ZZ}$ entangling gates. We separate the non-parametrized sector of the ansatz [$U_I$ in Eq.~(\ref{eq:ansatz})] from the parametrized part  [$U(\boldsymbol{\theta})$] with a dashed line, with the non-parametrized sector given by Hadamard gates $H = [\hat{\sigma}^{(x)}+\hat{\sigma}^{(z)}]/\sqrt{2}$.}
\label{hva_ansatz}
\end{figure}
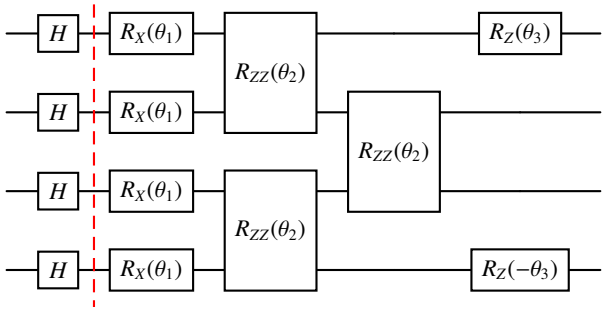

We start by describing the implementation of the VQE algorithm used to obtain the results in Fig.~\ref{fig:energy_magkink}. As mentioned in the previous section, we have employed the HVA ansatz in Eq.~(\ref{eq:circuit-unitary}) for preparing the trial states $\ket{\psi(\boldsymbol{\theta})}$. The associated HVA circuit for the Hamiltonian in Eq.~(\ref{eq:hamiltonian}) has the usual layered structure of entangling gates $R_{ZZ}(\theta) = \exp\left[(-i/2)\,\theta \,\hat{\sigma}^{(z)} \otimes \hat{\sigma}^{(z)}\right]$ between nearest neighbors in the spin lattice followed by rotations $R_X(\theta) = \exp\left[(-i/2)\,\theta \,\hat{\sigma}^{(x)}\right]$, as shown in Ref.~\cite{Wiersema:2020}, where we consider each spin in the lattice mapped to a single qubit of the circuit. Following the procedure adopted in Eq.~(\ref{eq:circuit-unitary}), to include the boundary field terms in the circuit (whose Hamiltonian coefficients are $h_l$ and $h_r$), we apply $R_Z(\theta) = \exp\left[(-i/2)\,\theta \,\hat{\sigma}^{(z)}\right]$ gates the qubits mapped to the spins at the ends of the chain, for every  layer of the ansatz. For the particular case of the magnet-to-kink transition ($h_l=-h_r$), we have considered these $R_Z$ rotations to have the same parameter and opposite directions, or $R_{Z1}(\phi)=R_{ZL}(-\phi)$, for some generic angle $\phi$, where $\{1,L\}$ represent the first and last qubits in the circuit layout. Considering this structure for the ansatz, we have a total of $3p$ parameters, where $p$ is the number of layers of the circuit. In Fig.~\ref{hva_ansatz}, we show a circuit representation, for a single layer ($p=1$), of the mentioned HVA ansatz for $L=4$. For the cases for $L=\{4,8,12\}$ shown in Figs.~\ref{fig:energy_magkink} and~\ref{fig:hardware_magkink_4q}, we have used $p=\{6,16,28\}$ layers, respectively.

For the optimization step of the VQE algorithm, the evaluations of the cost function in Eq.~(\ref{eq:cost-function}) were performed on a classical computer by considering the explicit representations of $\hat{H}$ and $\ket{\psi(\boldsymbol{\theta})}$, therefore disregarding the effects of both shot noise and hardware noise. In the first iteration of the quantum-classical feedback loop, the initial parameters $\theta_0$ were sampled uniformly at random in the range $(-0.3\pi,0.3\pi)$. For the following iterations, the values of $C(\boldsymbol{\theta})$ and $\boldsymbol{\theta}$ are given as input to the LBFGS-B optimization algorithm as implemented in SciPy \cite{scipy}. For computing the ground-state energies for various values of $h_l$, we have performed a random choice of initial parameters only for the first value of $h_l$ considered. For the subsequent values, we have used the optimal parameters found for the previous value of $h_l$ as starting point of the optimization. Since we expect the ground states of systems with similar values of $h_l$ to have a high overlap, this procedure can reduce the number of optimization iterations for different values of the Hamiltonian coefficients. For all the VQE optimizations shown in Fig.~\ref{fig:energy_magkink}, we have run a maximum of 7000 iterations of the LBFGS-B algorithm for the first value of $h_l$, and 1000 for the following values. After obtaining the optimal parameters $\theta_{\textbf{opt}}$ for all the considered values of $h_l$, we can compute the expectation value of any observable $\hat{O}$ on the ansatz state initialized with these parameters, or $\bra{\psi(\boldsymbol{\theta}_{\text{opt}})}\hat{O}\ket{\psi(\boldsymbol{\theta}_{\text{opt}})}$. For the results in Figs.~\ref{fig:energy_magkink} and~\ref{fig:hardware_magkink_4q}, we show the expectation values of $\hat{H}$ and $\hat{N}_k,$ which were computed using the optimal parameters obtained with VQE. We have used the Qiskit SDK \cite{qiskit2024} for all the calculations associated with building the circuits and computing expectation values. 

For the quantum hardware results shown in Fig.~\ref{fig:hardware_magkink_4q}, we have evaluated the expectation values of $\hat{H}$ and $\hat{\sigma}^{(i)}_z$ in the state prepared with the HVA ansatz initialized with the optimal parameters obtained with noiseless simulations of the VQE algorithm. 
 To obtain the 10 data points of the hardware results as in Fig.~\ref{fig:hardware_magkink_4q}, the value of each point is taken as the average for the energy expectation value over 15 separate runs. They were taken one after another, on the same day, according to the queue for the usage of the quantum processor, totaling 150 expectation values taken on the device. The number of shots for each run was 22304, and the run time for each run was 57s on average.
To access the $\texttt{ibm\_torino}$ processor via cloud, we have used the Qiskit Runtime plataform \cite{qiskit2024}. To mitigate hardware noise from the device, we have used the randomized compiling~\cite{Wallman:2016} and the twirled readout error extinction (T-REX)~\cite{VanDenBerg:2022} methods, with the standard configurations found in Qiskit IBM Runtime.

\section{Conclusion}
We have proposed a quantum simulation workflow for predicting the critical behavior of spins chains induced by local magnetic fields, including their boundary effects. By computing the ground-state energy of the system with the VQE algorithm for different values of local magnetic fields and sizes of the system, we were able to predict the critical values of the local  fields responsible for QPTs. If applied to larger system sizes in a physical quantum hardware, this approach has the potential to experimentally emulate boundary effects induced by local magnetic fields in a variety of critical spin systems. 

We have performed noiseless simulations of the VQE algorithm applied to spins chains of sizes $L=\{4,8,12\}$, where we were able to predict the ground-state energy, the corresponding derivatives, and also expectation values of observables evaluated on the ground state, with considerable accuracy when compared to the exact values. We have also shown that the circuit parameters obtained with VQE can also be used to prepare high-quality approximations of the ground state of the system on a quantum device, which in turn have been used to probe the corresponding critical behavior as well as other physical properties used to describe the system, such as the local magnetization. We have also shown that, with the approximate ground state prepared on the device, we are able to recover meaningful information on the system even in the presence of hardware noise, showing that our approach is well adapted to the limitations found on pre-fault tolerant devices. Our findings can also contribute to a deeper understanding of QPTs and their corresponding quantum simulations in larger and more complex systems. In particular, the convergence of VQE and its circuit complexity near criticality~\cite{SciPost:2023} can be explored for QPTs of different orders in realistic quantum computers, which can be addressed in terms of the scaling of the error function in the VQE algorithm. We leave these points for future investigations. 

{\section*{Acknowledgments}} 
We acknowledge the use of IBM Quantum services and IBM Quantum Credits for this work. The views expressed are those of the authors, and do not reflect the official policy or position of IBM or the IBM Quantum team.
We thank Alan C. Santos for useful discussions in the initial stage of this work. 
This study was financed in part by the Coordena\c{c}\~ao de Aperfei\c{c}oamento de Pessoal de N\'{\i}vel Superior - Brasil (CAPES) – Finance Code  001. 
A.D. acknowledges CAPES for financial support. 
M.S.S. is supported by Conselho Nacional de Desenvolvimento Cient\'{\i}fico e Tecnol\'ogico (CNPq) under the grant number 303836/2024-5. 
We also acknowledge the Brazilian National Institute for Science and Technology of Quantum Information (INCT-IQ).

\begin{figure}[!ht]
    \centering
    \includegraphics[scale=0.97]{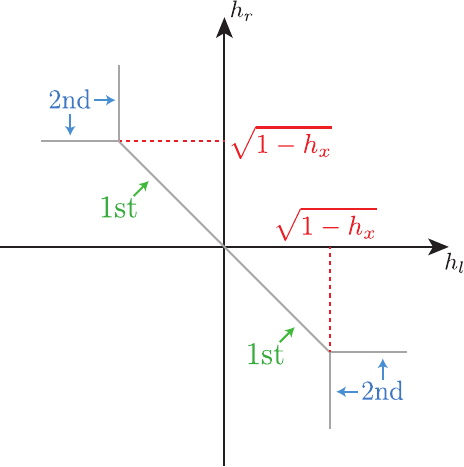}
    \caption{Quantum phase diagram of $\hat{H}$ in terms of the boundary fields, $h_l$ and $h_r$. The arrows in the figure indicate the boundaries corresponding to the first (1st) and second (2nd) order QPTs that occur in this system.}
    \label{fig:S1}
\end{figure}

\begin{figure*}[ht]
    \centering
    \includegraphics[width=15.5cm]{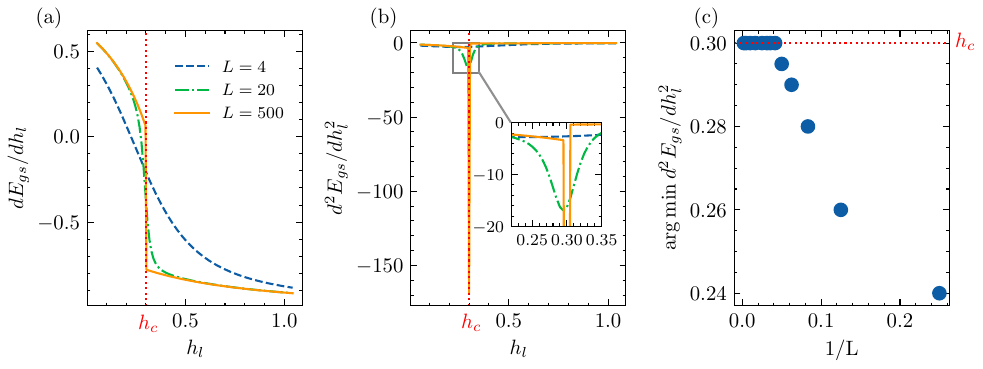} 
    \caption{Same as Fig.~\ref{fig:jordan_wigner_magkink} but considering $h_x = 0.8$, $h_r = -0.3$. By inspecting the indication of a discontinuity in the first derivative for $L=500$, we can characterize this transition as of first order. The plots are in units such that $J=1$.}
    \label{fig:S2}
\end{figure*}
\begin{figure*}[ht] 
    \centering 
    \includegraphics[width=15.5cm]{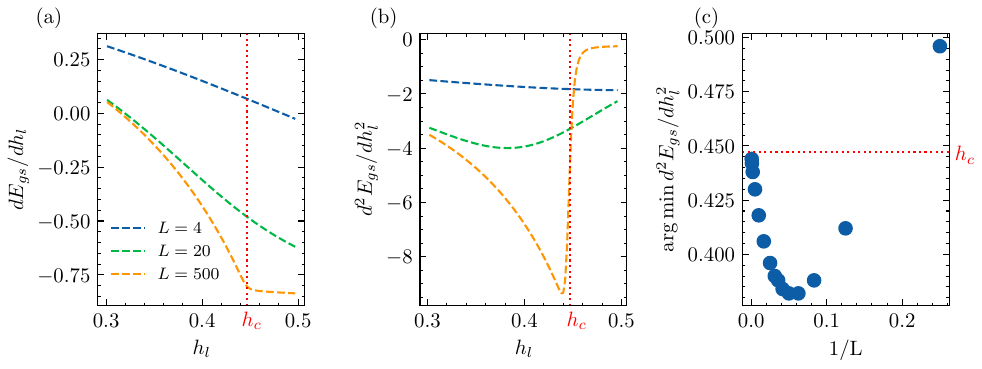} 
    \caption{Same as Fig.~\ref{fig:jordan_wigner_magkink} but considering $h_x = 0.8$, $h_r = 1.0$. By inspecting the indication of a discontinuity in the second derivative for $L=500$, we can characterize this transition as of second order. The plots are in units such that $J=1$.}
    \label{fig:S3}
\end{figure*}

\begin{figure*}[t] 
    \centering
    \includegraphics[width=15.5cm]{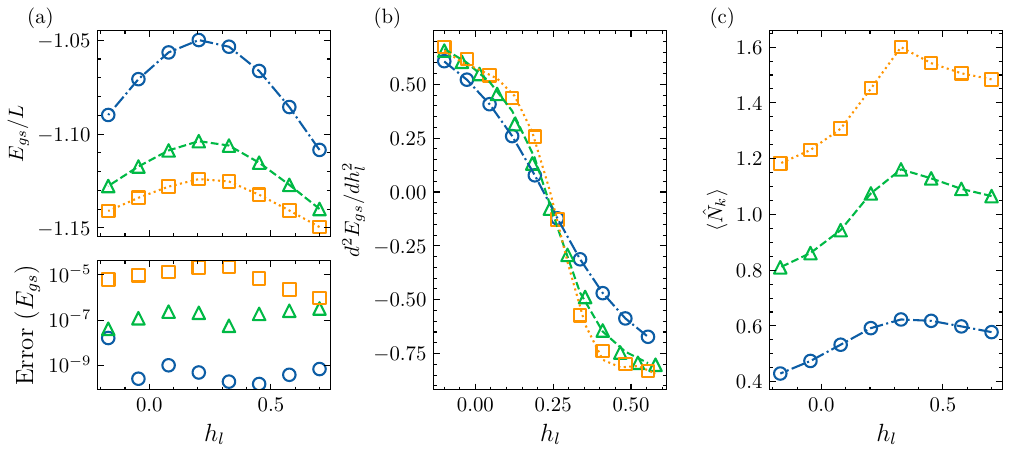} 
    \caption{Same as 
    Fig.~\ref{fig:energy_magkink} but considering $h_x = 0.8$, $h_r = -0.3$, and $h_l$ in the range $(-0.2,0.7)$. The plots are in units such that $J=1$.}
    \label{fig:S4}
\end{figure*}

\begin{figure*}[t] 
    \centering
    \includegraphics[width=15.5cm]{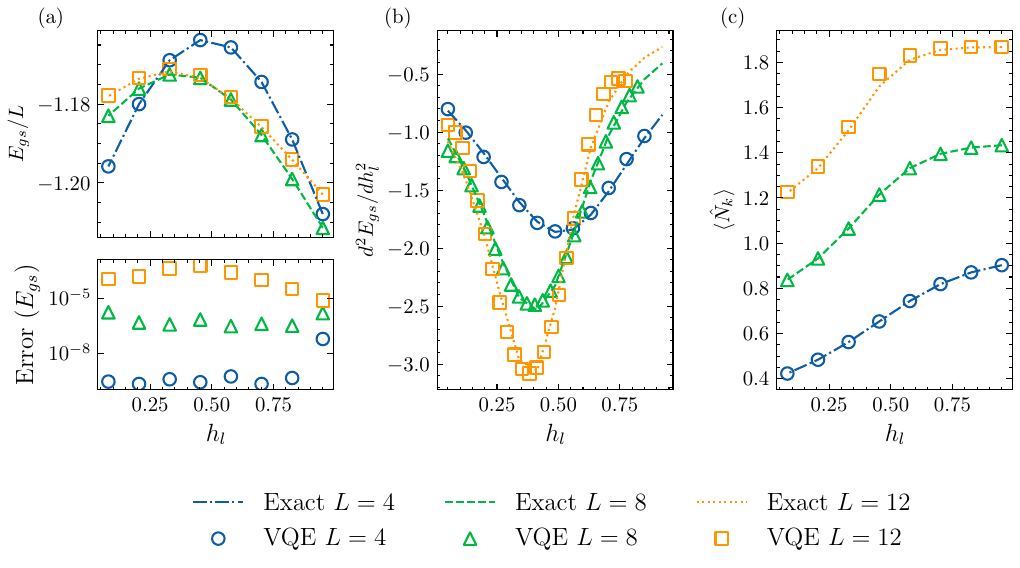} 
    \caption{Same as Fig.~\ref{fig:energy_magkink} but considering $h_x = 0.8$, $h_r = 1.0$, and $h_l$ in the range $(0.05,0.95)$. The plots are in units such that $J=1$.}
    \label{fig:S5}
\end{figure*}

\appendix 

\vspace{0.6 cm}
\section{Exact diagonalization of the transverse-field Ising chain with boundary magnetic fields}
\label{App_A}

Let us revisit the analytical treatment of the transverse-field Ising chain with boundary magnetic fields. For convenience, we rotate the spin axes by 90 degrees about the $y$ axis so that $\hat{\sigma}^x \rightarrow \hat{\sigma}^z$ and $\hat{\sigma}^z \rightarrow - \hat{\sigma}^x$. Then, the Hamiltonian provided by Eq.~(\ref{eq:hamiltonian}) becomes
\begin{align}
    \hat{H} = -J\sum_{i=1}^{L-1}\hat{\sigma}^{(x)}_i \hat{\sigma}^{(x)}_{i+1} -g\sum_{i=1}^{L}\hat{\sigma}^{(z)}_i-h_l \hat{\sigma}^{(x)}_1-h_r \hat{\sigma}^{(x)}_L ,\label{eq:hamiltonian2}
\end{align}
where we have taken $h_x\equiv g$. We can now replace the Hamtiltonian in Eq.~(\ref{eq:hamiltonian2}) for an effective Hamiltonian $\hat{H}_{eff}$, where we append one additional spin to the left and right sides of the original chain, labeled by the indices $0$ and $L+1$, respectively. 
Then, the corresponding Hamiltonian reads
\begin{eqnarray}
    \hat{H}_{eff} &=& -J\sum_{i=1}^{L-1}\hat{\sigma}^{(x)}_i \hat{\sigma}^{(x)}_{i+1} -g\sum_{i=1}^{L}\hat{\sigma}^{(z)}_i \nonumber \\
    &&-\left|h_l\right| \hat{\sigma}^{(x)}_0 \hat{\sigma}^{(x)}_1-\left|h_r \right|\hat{\sigma}^{(x)}_L \hat{\sigma}^{(x)}_{L+1} . \label{eq:hamilt_eff}
\end{eqnarray}
Notice that, since the transverse field does not affect the boundary spins $0$ and $L+1$, we have that $\left[\hat{\sigma}^{(x)}_0,\hat{H}_{eff}\right]=0$ and $\left[\hat{\sigma}^{(x)}_{L+1},\hat{H}_{eff}\right]=0$. This means that the operators $\hat{\sigma}^{(x)}_{0}$, $\hat{\sigma}^{(x)}_{L+1}$, and $\hat{H}_{eff}$ can be simultaneously diagonalized, with the Hilbert space divided into four sectors. Those sectors correspond to the eigenvalue pairs $(s_0,s_{L+1})$ for $\hat{\sigma}^{(x)}_{0}$ and $\hat{\sigma}^{(x)}_{L+1}$, which are labeled by $(1,1)$, $(1,-1)$, $(-1,1)$, and $(-1,-1)$. 

In order to discuss the quantum phase transitions of the original chain, we will consider the energy  spectrum for anti-parallel fields $h_L$ and $h_R$. More specifically, we will take the configuration $h_L>0$ and $h_R<0$, which is associated to the sector $(1,-1)$ of the extended Hilbert space. The opposite case,  $h_L<0$ and $h_R>0$, can be directly obtained by symmetry. 
The exact diagonalization of the Hamiltonian follows by mapping the Hilbert space of spin$-1/2$ particles into the Hilbert space of spinless fermions hopping between sites with single orbitals. This is implemented through the Jordan-Wigner transformations
\begin{eqnarray}
\hat{\sigma}^z_i &=& 1 - 2 \hat{c}^\dagger_i \hat{c}_i, \nonumber \\
\hat{\sigma}^x_i &=&  \prod_{j<i} \left(1-2\hat{c}^\dagger_j \hat{c}_j\right)\left(\hat{c}_i+\hat{c}_i^\dagger\right),
\label{jw}
\end{eqnarray}
where $\hat{c}_i^\dagger$ and $\hat{c}_i$ are the creation and annihilation fermion operators at site $i$, respectively, obeying the anticommutation relations 
$\left\{\hat{c}_i^\dagger,\hat{c}_j\right\}=\delta_{i,j}$, 
$\left\{\hat{c}_i^\dagger,\hat{c}_j^\dagger\right\}=0$, and $\left\{\hat{c}_i,\hat{c}_j\right\}=0$. Therefore, rewriting the model in terms of the fermion variables, we obtain the quadratic Hamiltonian
\begin{align}
    \hat{H}_{eff} = \sum_{i,j=0}^{L+1}
    \left[ \hat{c}_i^\dagger A_{ij} \hat{c}_j + 
    \frac{1}{2} \hat{c}_i^\dagger B_{ij} \hat{c}_j^\dagger - 
    \frac{1}{2} \hat{c}_i B_{ij} \hat{c}_j \right] - g L, \label{eq:hamilt_eff_fermion}
\end{align}
with 
\begin{eqnarray}
    &&A_{ij} =  \left[ -J (1-\delta_{i,0})
    (1-\delta_{i,L+1})(1-\delta_{j,0})
    (1-\delta_{j,L+1}) \right. \nonumber \\ 
    &&\left. - \left| h_L \right| (\delta_{i,0}+\delta_{j,0}) -\left| h_R \right| (\delta_{i,L+1}+\delta_{j,L+1})\right] (\delta_{j,i+1}+\delta_{i,j+1}) \nonumber \\ 
    && - 2g (1-\delta_{i,0})
    (1-\delta_{i,L+1}) \delta_{i,j} 
     \label{eq:Aij}
\end{eqnarray}
and 
\begin{eqnarray}
    &&B_{ij} =  \left[ -J (1-\delta_{i,0})
    (1-\delta_{i,L+1})(1-\delta_{j,0})
    (1-\delta_{j,L+1}) \right. \nonumber \\ 
    &&\left. - \left| h_L \right| (\delta_{i,0}+\delta_{j,0}) -\left| h_R \right| (\delta_{i,L+1}+\delta_{j,L+1})\right] (\delta_{j,i+1}-\delta_{i,j+1}). \nonumber \\
     \label{eq:Bij}
\end{eqnarray}

Now we can complete the diagonalization by performing the Bogoliubov transformation
\begin{equation}
\eta_k = \sum_{i=0}^{L+1} \left( g_{ki} \hat{c}_i^\dagger + h_{ki} \hat{c}_i \right),
\end{equation}
where $g_{ki}$ and $h_{ki}$ are determined by imposing the requirement that the effective Hamiltonian takes the diagonal form
\begin{equation}
\hat{H}_{eff} = E_{gs} + \sum_{k=0}^{L+1} \varepsilon_k \hat{\eta}_k^\dagger \hat{\eta}_k,
\end{equation}
with $E_{gs}$ the ground state energy and $0\le \varepsilon_1\le \varepsilon_2 \le \cdots$. Following Ref.~\cite{Campostrini:15JSTAT}, we can evaluate $E_{gs}$ by imposing
\begin{equation}
    E_{gs}=-\frac{1}{2} \sum_{k=0}^{L+1} \varepsilon_k,
    \label{Egs}
\end{equation}
where $\varepsilon_k$ can be obtained by solving the eigenvalue problem
\begin{equation}
(A+B)(A-B) |\varphi_k\rangle = \varepsilon_k^2 |\varphi_k\rangle 
\end{equation}
with $A$ and $B$ denoting the matrices defined by Eqs.~(\ref{eq:Aij}) and (\ref{eq:Bij}), 
respectively. In the sector $(1,-1)$ of the extended Hilbert space, the lowest energy 
state is the first excited state $\eta_1^\dagger |0\rangle$ of the Hamiltonian $\hat{H}_{\textrm{eff}}$~\cite{Campostrini:15JSTAT}.

From the exact expression for the lowest energy state in the sector $(1,-1)$, we can derive the quantum phase diagram for the model. This is illustrated in Fig.~\ref{fig:S1} (adapted from Ref.~\cite{Hu:2021}). 
This phase diagram shows different kinds of QPTs in the Hilbert sectors $(1,-1)$ and $(-1,1)$. Specifically, we can see that, by following a path on the $h_l\times h_r$ plane, there will be a QPT if the path crosses the indicated phase boundaries. If it crosses the vertical and horizontal lines, the corresponding QPT will be continuous of second-order, with the critical value depending on the intensity of the transverse field $h_x$ (taken as $g$ in this Appendix). On the other hand, the diagonal boundary is associated with a first-order QPT with a boundary following the curve $h_r = -h_l$.

\begin{figure*}[t] 
    \centering
    \includegraphics[width=15.5cm]{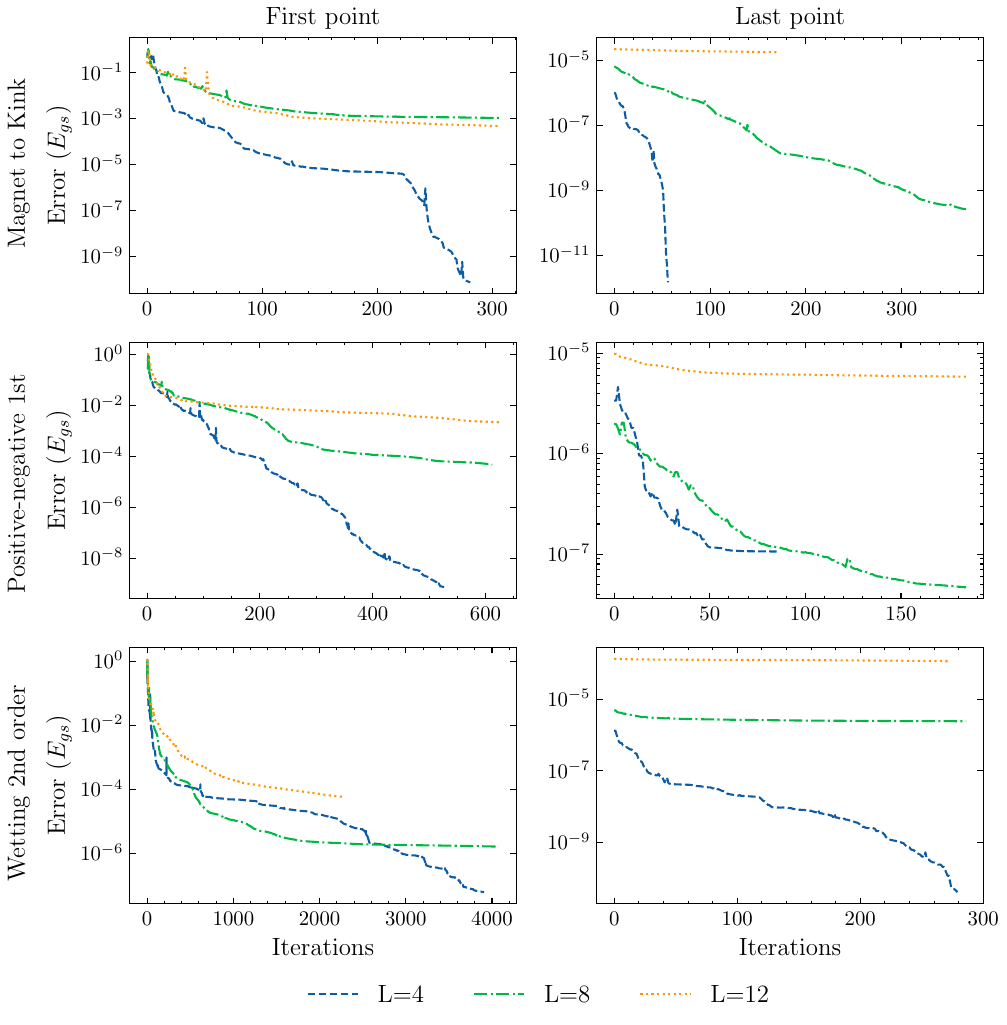} 
    \caption{Error in the prediction of the ground state energy along the iterations of the LBFGS-B algorithm, for all the phase transitions described in this work. We highlight how, in the case of the last point, in which the initial parameters for the optimization were recycled from the previous value of $h_l$, the error in the first iterations is considerably small, in comparison with the first point in which the initial parameters were sampled at random.}
    \label{fig:S6}
\end{figure*}

\section{VQE results for first-order and second-order QPTs }

\label{App_B}

Let us present now the results for the quantum critical point, but considering other configurations of the transverse-field Ising Hamiltonian with boundary magnetic fields. 
Considering the possible trajectories  in the quantum phase diagram, the QPT discussed in the main text is associated with a particular second-order QPT, where the trajectory coincides with the boundary of the first-order QPT. Here we compare this particular magnet-to-kink behavior with other possible transitions. 
In Fig.~\ref{fig:S2} and Fig.~\ref{fig:S3}, we show the exact results for the derivatives of the ground state energies (similar as Fig.~\ref{fig:jordan_wigner_magkink}), considering a first-order QPT and also a second-order QPT, with a different trajectory from the magnet-to-kink critical line. By inspecting the results shown for $L=\{4,20,500\}$, we can infer that, in the limit $L\rightarrow\infty$, the curve of the first derivative of the ground state energy will become discontinuous at the predicted first-order critical point, while in the second-order case, a non-analyticity is present in the second derivative of the ground-state energy.

We also show how the argument of the minimum of the second derivative approaches the critical field as $L\rightarrow\infty$. By comparing this result with Fig.~\ref{fig:jordan_wigner_magkink}, we can identify the magnet-to-kink transition shown in Ref.~\cite{Campostrini:15} as being of second order, by identifying the non-analyticity only in the second derivative of the ground state energy.
The results obtained with VQE can  also be discussed. In Fig.~\ref{fig:S4} and Fig.~\ref{fig:S5}, we show the same analysis shown in Fig.~\ref{fig:energy_magkink} applied for these other paths in the quantum phase diagram. Moreover, in Fig.~\ref{fig:S6}, we show the error for the ground state energy given by VQE along the iterations of the optimization, for the first and last values of the boundary field $h_l$. We  refer to the transition shown in the main text as "magnet to kink", and the first and second order transitions shown in this appendix as "positive-negative 1st" and "wetting 2nd order", respectively.
We highlight how the optimization starts in a much smaller error in the first iteration for the case of the last value of $h_l$, and that is associated with the choice of using the initial parameters of the optimization as the ones coming from the previous value of $h_l$. We have taken the first value of $h_l$ as the largest one, iterating over this parameter in decreasing order. Then, the first point indicated refers to the largest value of $h_l$, while the last point refers to its smallest value. 

\section{Root mean square deviation in energy}

\label{app_C}

 As shown in Fig.~\ref{fig:hardware_magkink_4q}(a), the error in energy is small as the boundary field is varied. However, simulation on noisy hardware induces significant changes in the derivatives of energy. Indeed, we are interpolating a noisy energy curve, which leads to an amplified fluctuation in the energy derivatives. This is due to the very nature of the interpolation process. The statistical error for the ground state energy can be measured by the root mean square (RMS) deviation of data samples, comparing the energy and its derivatives obtained by exact results and by the actual quantum hardware with the noiseless VQE results. The RMS error for a sample of $n$ points 
$\{x_i\}$ with respect to $\{y_i\}$, with $1\le i \le n$, can be defined as 
\begin{equation}
RMS(x,y) = \sqrt{\frac{1}{n}\sum_{i=1}^{n}(x_i-y_i)^2}. 
\label{RMSxy}
\end{equation}
We can then use Eq.~(\ref{RMSxy}) to compare a set of exact energy results $\{E_i\}$ and a set of actual quantum hardware values 
$\{H_i\}$ with the corresponding noiseless VQE results 
$\{V_i\}$, with the lists $\{E_i\}$, $\{H_i\}$, and $\{V_i\}$ obtained from the energy points for a fixed range of values for boundary fields $h_l$. By taking $h_l$ as the experimental points exhibited in Fig.~\ref{fig:hardware_magkink_4q}(a), the results for the RMS error are then  
$RMS(H,V)=0.28$ and $RMS(E, V) = 0.0030$. Notice that $RMS(H,V)$ is much larger than $RMS(E, V)$. This suggests that most of the statistical error in the experimental data comes from the hardware noise compared to the error in the noiseless VQE approximation for a finite number of layers. If we now compute the statistical error in the second derivative of the energy, considering the samples $\{dE_i\}$, $\{dH_i\}$, and $\{dV_i\}$ obtained from the interpolation functions by exact results, actual hardware, and noiseless VQE, respectively, we obtain $RMS(dH,dV)=0.53$ and $RMS(dE, dV) = 0.053$. From these results, we can again see that the overall error due to hardware noise dominates over the algorithmic error. Moreover, both the errors $RMS(dH,dV)$ and $RMS(dE, dV)$ are considerably larger than their counterparts $RMS(H,V)$ and $RMS(E, V)$. This increase in error can be attributed to the interpolation procedure itself, since the interpolated curve originated from noisy data samples.

\end{document}